\documentclass[aps,prb,preprint,showpacs]{revtex4} 
\usepackage{graphicx}
\usepackage{hhline}
\usepackage{color}
\usepackage{amsmath}

\begin{document}

\title{Reinvestigation on large perpendicular magnetic anisotropy in Fe/MgO interface from first-principles approach}

\author{Nurul Ikhsan,$^{1,2}$ Tomosato Kanagawa,$^{1}$ Indra Pardede,$^{1,3}$\\
 Masao Obata,$^{1,4}$ and Tatsuki Oda$^{1,4}$}

\affiliation{
$^{1}$Graduate School of Natural Science and Technology, Kanazawa University, Kakuma, Kanazawa 920-1192, Japan\\
$^{2}$School of Computing, Telkom University, Bandung 40257, Indonesia\\
$^{3}$Department of Physics, Institut Teknologi Sumatera, Lampung Selatan 35365, Indonesia\\
$^{4}$Institute of Science and Engineering, Kanazawa University, Kakuma, Kanazawa 920-1192, Japan}


\begin{abstract}
We investigated electronic structure and magnetic anisotropy in the Fe/MgO interface of magnetic metal and dielectric insulator under the Cr layer of small electronegativity, by means of the first-principles density functional approach. 
The result indicates that the interface resonance state gets occupied unlike a typical rigid band picture as the number of Fe layers decreases, finding large perpendicular anisotropies in the oscillating behavior for thickness dependence. 
We discuss scenarios of the two dimensional van Hove singularity associated with flat band dispersions, and also the accuracies of anisotropy energy in comparison with the available experimental data. 
\end{abstract}


\maketitle



Interface between magnetic metal and insulator has been attractive for a long time in material physics.
In particular, the interface of Fe/MgO or its family play a key role in the magnetic tunnel junction for the development of magnetoresistive random access memory (MRAM), being used in the typical high performance system of Fe/MgO/Fe junctions.\cite{Yuasa2004, Ando2005, Butler2001, Koo2014} 
These devices, currently driven by the mechanism of spin transfer torque,\cite{Slonczewski1996, Berger1996} require a large ratio of tunnel magnetoresistances for MRAM applications. 

Surprisingly, such interface has been also investigated for development of the possible MRAM which works with electric-field (EF) magnetization reversal mechanism. 
 Such multi-functional interface has a lot of interesting physical properties and potential applications for magnetic and magnonic devices.\cite{Kruglyak2010}
Its unique magnetic characteristic, perpendicular magnetic anisotropy, voltage/spin-current controlled magnetic anisotropy, also unique spin structure with voltage controlled Dzyaloshinskii-Moriya interaction.\cite{Nawaoka2015APEX}
These features are promising for information technology device, especially for high density and low power consumption data storage.\cite{ShiotaNM2012} 

In solid state physics, the two dimensional electronic structures originated by the van Hove singularity (vHS) of band theory has provided us many interesting physics as well as those of other (one and three) dimensional  properties; in carbon nanotubes the optical transition between the vHS states,\cite{CarbonNT} the high-$T_{\rm c}$ H$_{x}$S superconductors, etc.\cite{Sano2016}
In the works for development of the devices, the family of Fe/MgO interfaces has been used as a kernel technological element.
They have showed a strong perpendicular magnetic anisotropy for the thin Fe layer without any heavy element.\cite{Ikeda2010, Shimabukuro2010} 
The multi-functional properties, mentioned in the previous paragraphs, are mostly originated from the electronic structure.
In such system, the interface resonance state (IRS) has been discussed in the several works.\cite{Zermatten2008}
These states are consequences of the band formation consisting of non-bonding orbitals on the interface.
Although such character has been observed as interesting characteristics, the van Hove's singularity(vHS) has never been pointed out so far in terms of orbital components at the interface. 

In recent improvement of computational performance allows us to estimate the magnetic anisotropy or its EF effect precisely and numerically.\cite{Duan2008, Nakamura2009} Such improvement contributes not only to physical and qualitative explanations in the property of magnetic anisotropy, but also to semi-quantitative agreements. In particular, the slope in the EF variation has been proved to have a realistic meaning, when compared with the experimental results.\cite{Nozaki2010} On the magnetic anisotropy energy (MAE) itself, the comparison with experiment has a distance from explaining the experimental measurements with a quantitative agreement. The experimental progresses on the interface magnetic anisotropy in the thin films give us a fascinating opportunity on a direct comparison between the theoretical and experimental approaches. 

This work was devoted to the discussions on electronic, magnetic, and structural properties of Fe/MgO(001), as a reinvestigation in the view point of two dimensional vHS. We obtained remarkable Fe-thickness dependences of MAE, implying a picture of non-rigid band filling in the IRS. Such result can be discussed in terms of electronic band theory, compared with the available experimental data. 
In the comparison, we also discuss the computational accuracy on MAE within the current state-of-the-art for density functional theory (DFT). 
 
\begin{figure}[t]
\begin{center}
\includegraphics[width=85mm]{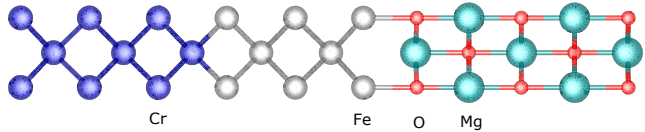}
\end{center}
\caption{Schematic diagram of the slab systems; Cr(6ML)/Fe({\it x}ML)/MgO(5ML) for $x=5$.}
\label{FigSystem}
\end{figure}


In order to investigate the interface of Fe/MgO, we consider a slab system, vacuum (0.79 nm)/Cr($w$ ML)/Fe ($x$ ML)/MgO (5 ML)/vacuum (0.79 nm) (ML$=$atomic monolayer), in the computation (Fig. \ref{FigSystem}).
The atoms are specified by number as Cr(1), Cr(2), ..., Fe (1), Fe (2), ..., O (1), ..., Mg (1), ... etc. from the Fe/MgO interface.
At this interface an Fe atom was placed just next to the O atom due to its stability, and in the Cr and Fe layers the body-centered layer-stacking sequence was used.\cite{Parkin90,Johnson2007}
The in-plane lattice constant extracted from bulk Cr was employed.
The thickness of Fe thin layer ($x$) was varied from 1 ML until 10 ML, while that of Cr ($w$) was kept at $w=6$. The latter thickness is enough to give an constant MAE within an accuracy of 0.1mJ/m$^2$ for a given Fe thickness. 

We carried out first-principles density functional calculations, which employ fully relativistic (with spin-orbit interaction) and scalar-relativistic (without spin-orbit interaction) Ultrasoft pseudopotentials and planewave basis,\cite{Oda2005, Laasonen1993} by using the generalized gradient approximation.\cite{Perdew1992}
The MAE originated from spin-orbit interaction (SOI) was estimated from the total energy difference between the different magnetization directions [100] ($x$-axis) and [001] ($z$-axis), ${\rm MAE(SOI)}=E[100]-E[001]$, where [001] specifies the direction of film thickness (right-hand side in Fig. \ref{FigSystem}). 
We used the 32 $\times$ 32 $\times$ 1 mesh in {\bf k} point sampling\cite{Monkhost76} in MAE estimation.\cite{Tsujikawa2008}
Using the scalar-relativistic level computation, in which taking a 24 $\times$ 24 $\times$ 1 {\bf k}-mesh, we induced structural relaxation while keeping both the in-plane lattice constant and the atomic coordinates of O(3).
The MAE from the shape anisotropy was estimated using the magnetostatic dipole interaction (MDI) and assuming the atomic magnetic moments.\cite{Szunyogh1995} 

\begin{figure}[t]
\begin{center}
\includegraphics[width=85mm]{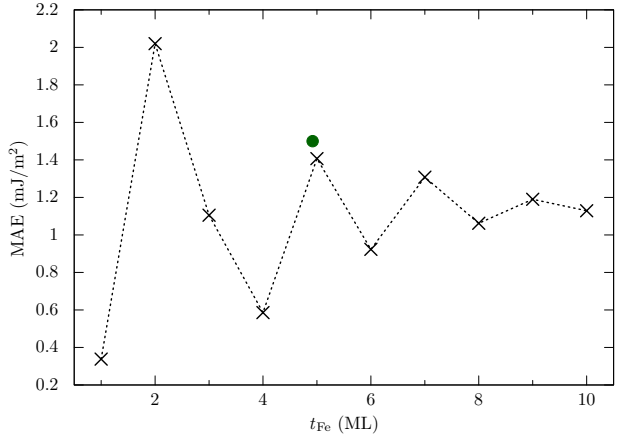}
\end{center}
\caption{Thickness dependence of the magnetic anisotropy energy (MAE) from spin-orbit interaction in Cr/Fe({\it x}ML)/MgO. The bullet indicates the experimental value (ref. \onlinecite{Koo2014}), where 1ML thickness is assumed to be 0.142 nm.}
\label{MAEenergy}
\end{figure}

We report the thickness dependence of MAE(SOI) in Fig. \ref{MAEenergy}. It indicates an oscillating perpendicular anisotropy with respect to Fe thickness ($x$), and the maximum of 2.0 mJ/m$^{2}$ at Fe 2ML and the maximal values at 5ML and 7ML. The behavior shows that an odd-even alternating oscillation at the thicker systems (4ML-10ML).  For the thinner systems, the amplitude of oscillation is largely enhanced since large changes are expected in the electronic structure around the Fe/MgO interface. The maximum MAE at Fe 2ML is much larger than the previous theoretical and experimental values in the Fe/MgO interface family, comparable to the interface contribution extracted from the extrapolation fitting in the experiment.\cite{NozakiPRA2016} At Fe 5ML our value agrees well with the experimental value.\cite{Koo2014} 

\begin{figure*}[t]
\hspace*{-5mm}
\begin{minipage}{8.0cm}
\includegraphics[width=70mm]{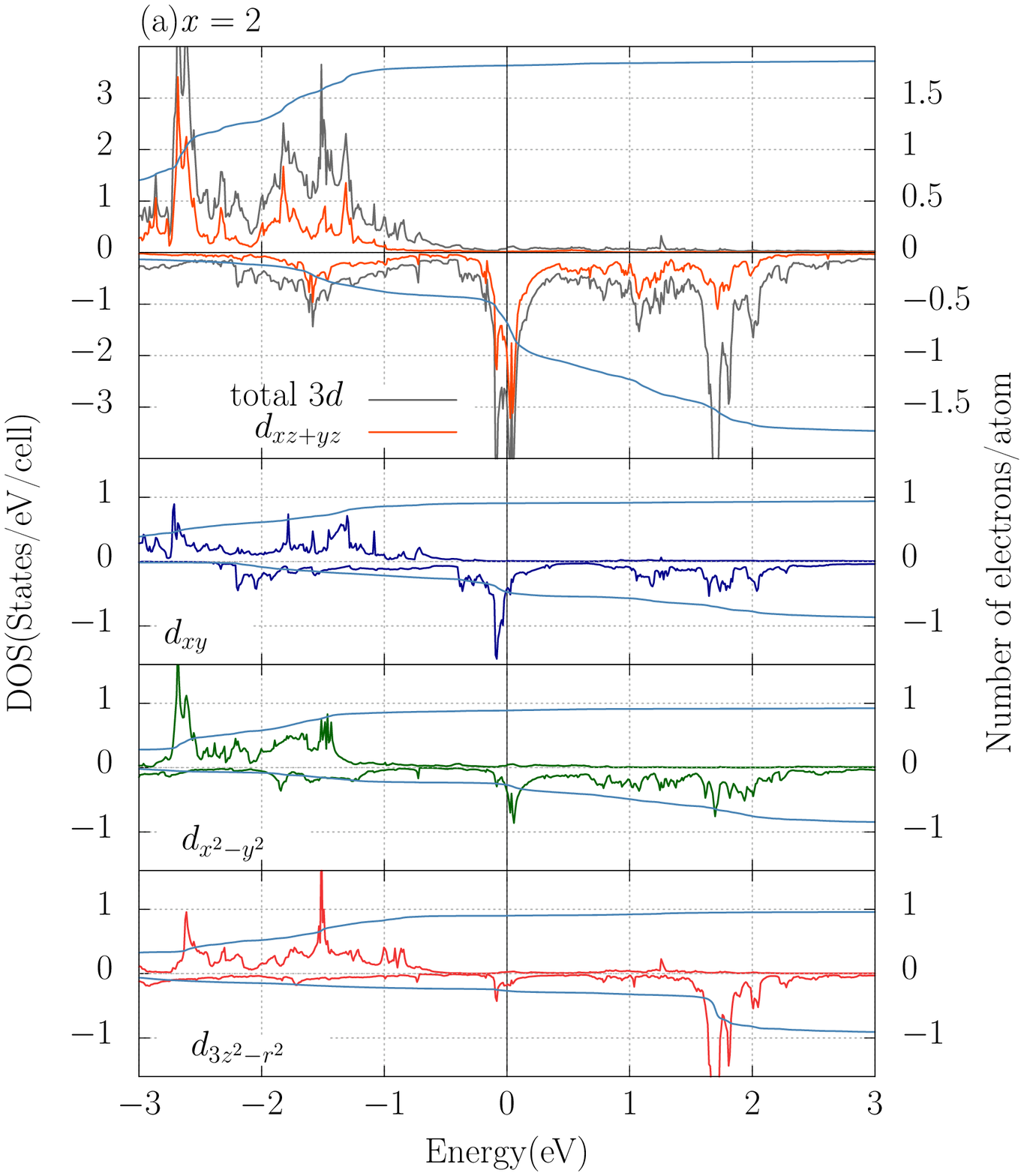}
\includegraphics[width=40mm]{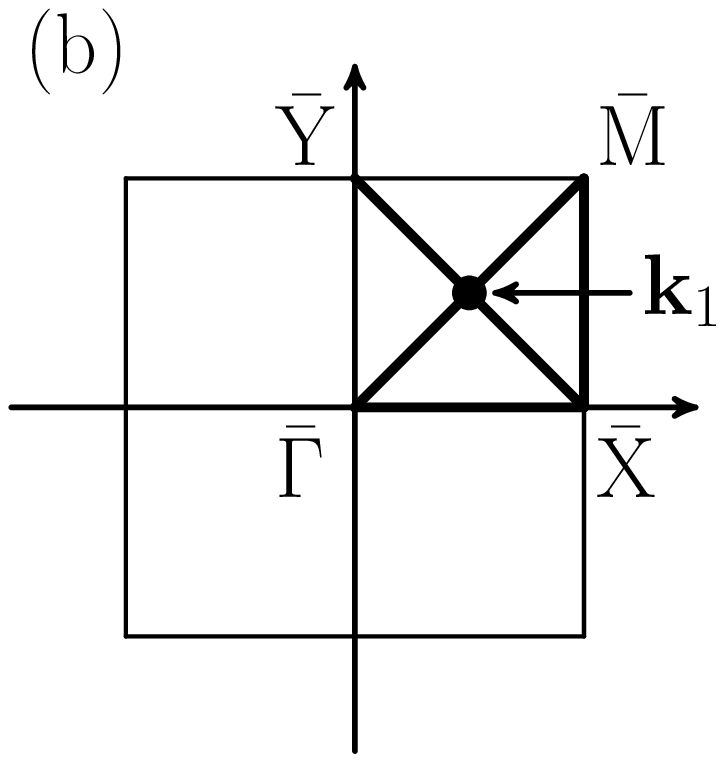}
\end{minipage}
\hspace*{-5mm}
\begin{minipage}{8.0cm}
\includegraphics[width=78mm]{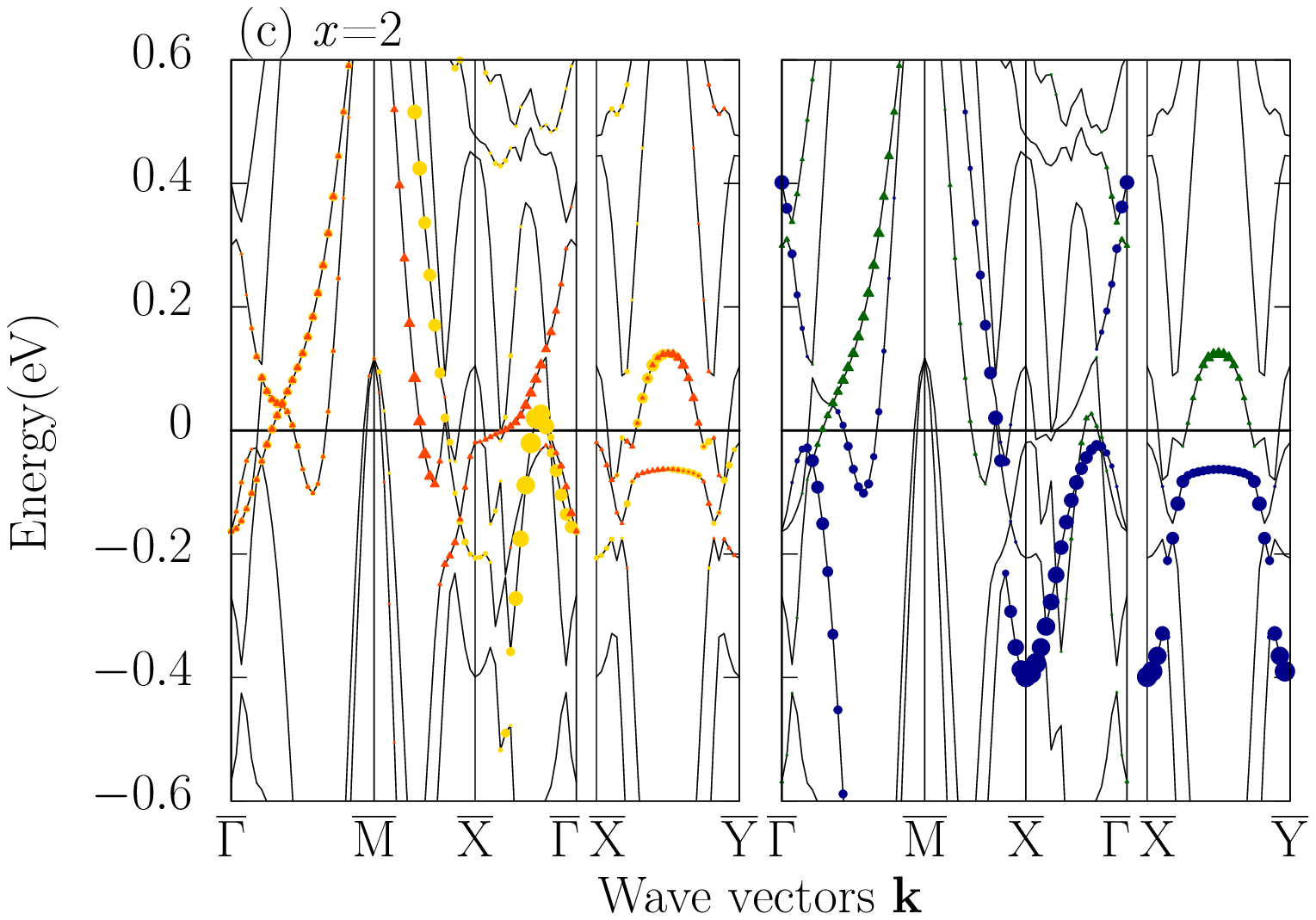}
\\[1mm]
\includegraphics[width=78mm]{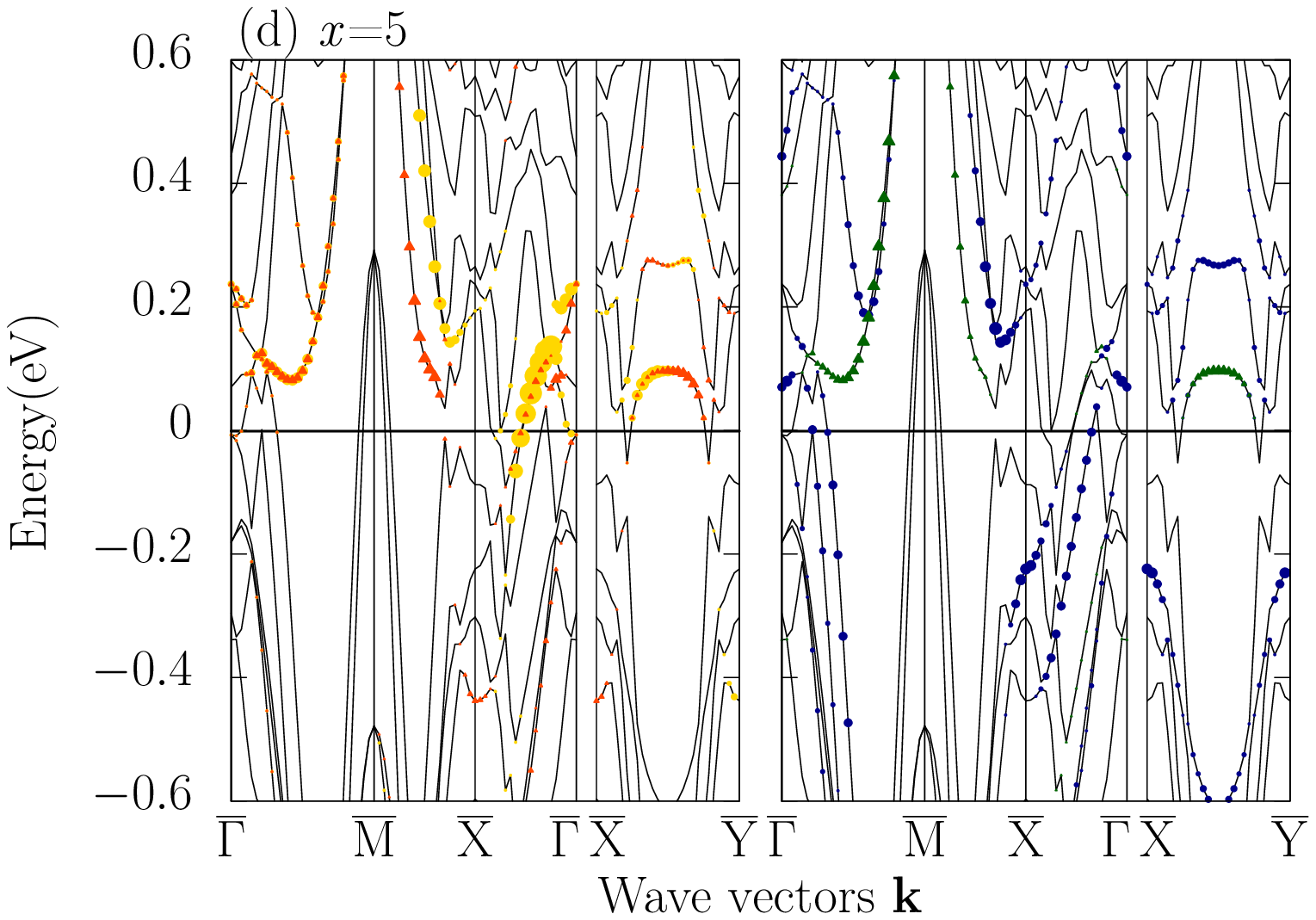}
\end{minipage}
\caption{Electronic structures at the Fermi level in Cr/Fe($x$ML)/MgO ($x=2, 5$) without including spin-orbit interaction ; (a) projected densities of states (PDOS) for the Fe ${\rm 3}d$ orbitals at the interface Fe(1), (b) 1st Brillouin zone, (c) band dispersions for $x=2$, (d) band dispersions for $x=5$. The colors of grey, orange (yellow), blue, green, and red are used for indicating the components of total 3$d$ orbital, $d_{xz+yz}$, $d_{xy}$, $d_{x^{2}-y^{2}}$, and $d_{3z^{2}-r^{2}}$, respectively.}
\label{PDOS-625}
\end{figure*}

Figure \ref{PDOS-625} shows the projected density of states (PDOS) at the interface in $x=2$ and band dispersions with 3d orbital components in the vicinity of the Fermi level ($E_{\rm F}$).
As shown, the $E_{\rm F}$ is located between the two peaks of PDOS consisting of 3d orbitals.
These electronic states forms a flat band around the {\bf k}-point ${\bf k}_{1}=\pi/a (1/2, 1/2)$ (see Fig. \ref{PDOS-625}) and a vHS in two-dimensional Brillouin zone (2DBZ).
The vHS is more clearly shown in $x=5$.
The band flatness remarkably appears on $\bar{\rm X}$--$\bar{\rm Y}$ line in 2DBZ and the saddle point nature appears near/around the cross point ${\bf k}_{1}$ of $\bar{\rm X}$--$\bar{\rm Y}$ and $\bar{\Gamma}$--$\bar{\rm M}$ lines.
These features are the origins of sharp PDOS peaks in the IRS, appearing more or less in the Fe/MgO and its family systems.
However, for realizing these features, there may be two conditions.
The one is an appropriate orbital hybridization between 1st Fe and 2nd Fe layers.
This keeps splitting the mixed eigenstates of $d_{xz}$ and $d_{yz}$ components at $\bar{\rm M}$ point to the lower and higher eigen energies, while in the Fe 1ML system, those stays remain on or around $E_{\rm F}$.\cite{WangPRB1993}
The 2nd condition is also an orbital hybridization between Fe $d_{3z^{2}-r^{2}}$ and O $p_{z}$.
This keeps the $d_{3z^{2}-r^{2}}$ away from $E_{\rm F}$, not disturbing at $E_{\rm F}$. 
The latter has been well known as one of important origins for realizing perpendicular anisotropy.\cite{Shimabukuro2010}
This is because the orbital $d_{3z^{2}-r^{2}}$ always contributes only to in-plane magnetic anisotropy, assuming that the contribution from the majority spin state can be neglected due to a large exchange splitting.\cite{WangPRB1993}
The MAE maximum in Fe 2ML is obtained as the consequences of the origins discussed above. Note that there are vertical couplings of SOI around ${\bf k}_{1}$ area in 2DBZ, which contribute to perpendicular anisotropy; couplings of $d_{xy}$--$d_{x^{2}-y^{2}}$ and $d_{xz}$--$d_{yz}$. 

The thickness dependence in MAE has a relationship with the number of electrons (NOE) in the 3d orbital of minority spin state on the interface Fe.
In the large variation range of MAE ($<$ Fe 4ML), the NOE decreases as thickness, and in the odd-even alternating range ($\ge$ Fe 4ML) its NOE does not synchronize so much with the MAE, as shown in the supplemental material (Fig. S1).
For the former, the number of Fe works as hole doping, while in the latter the MAE may be influenced by the details of electronic structures at the interface.
Interestingly, note that the NOEs of $d_{xy}$ and $d_{x^{2}-y^{2}}$ show an alternating behavior in their relative values (See Fig. S1). 
Indeed, the alternating nature appears at the $d_{xy}$ component in the band dispersions and PDOS (not shown).
The Fe layer, forming a quantum well structure, is terminated at the Fe/MgO and Cr/Fe interfaces, calling the amplitudes of wave function at the edges have a feature of odd-even alternating behavior in the one dimensional model chain of which the atomic site is connected with electron transfer integral.
The $d_{xy}$ component at the Fe/MgO interface is largely affected due to its non-bonding nature, and can be sensitive to external perturbation from the other edge of Cr/Fe interface.
The electron transfer between the $d_{xy}$ orbitals of neighboring Fe MLs is proportional to $T_{xy,xy}({\bf k}) \sim e^{i{\bf k}\cdot {\bf R}} \{ 3t(dd\sigma)+2t(dd\pi)+4t(dd\delta)\}/9$, and is not negligible because the absolute of transfer integral can reach to about 0.1 eV.\cite{WangPRB1993}  
Indeed, comparing the band dispersions between odd and even Fe MLs, the behaviors of $d_{xy}$ component are much different, in particular, around the $\bar{\rm X}$ point in 2DBZ at $E_{\rm F}$. 

\begin{figure}[t]
\begin{center}
\includegraphics[width=85mm]{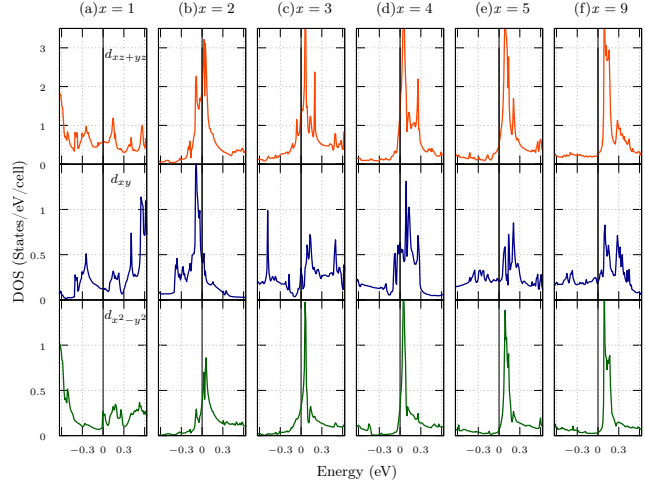}
\end{center}
\caption{Projected density of states (PDOS) for the interface resonance state in Cr/Fe({\it x}ML)/MgO; (a)--(f) for $x=1-5$ and 9, respectively. The top, middle, bottom lows in the panels show the components of $d_{xz+yz}$, $d_{xy}$, and $d_{x^{2}-y^{2}}$.}
\label{FigsetofPDOS}
\end{figure}

Figure \ref{FigsetofPDOS} shows the series of PDOS for IRS.
Except for $x=1$, the sharp peaks appear, implying the localized nature of wave functions.
It is worthy to notice that the series of $x=2-5$ do not show a simple rigid band filling.
The shape of $d_{xy}$ component changes while that of $d_{x^{2}-y^{2}}$ is kept without any large change.
In details, the peaks of  $d_{xy}$ and $d_{x^{2}-y^{2}}$ are located just below and above the $E_{\rm F}$ level, respectively, in Fe 2ML, and as the thickness the peaks of $d_{xz+yz}$ moves to higher energies with that of $d_{x^{2}-y^{2}}$.
Additionally, it is interesting to see the PDOS peak sharpen in the thicker systems (see the case of $x=9$).  
The origin of IRS energy shift, showing a property of smaller electronegativity for Cr, is speculated as an orbital hybridization with the Cr underlayer.
It was observed clearly in the Fe layer with small $x$'s as a vicinity effect of Cr, and such shift has been discussed in the literatures.\cite{Ogura2011}
Indeed, the Cr-orbital component also appears at the IRS. 
Interestingly, the similar energy lowering of IRS occurs as the decrease of in-plane lattice constant,\cite{Yoshikawa2013} inducing the modulations in MAE. 


Decreasing the thickness of Fe layer to, for example, 2ML, the ferromagnetic state may become unstable against thermal disturbance. This could be denied in a discussion of Cr--Fe magnetic interaction. The Fe at the Cr/Fe interface has a strong antiferromagnetic interaction of Heisenberg type. This stabilizes the ferromagnetic state of Fe thin layer. 
The previous works reported large exchange interactions of $J= 69$ meV and $|J|$=59, 20 meV for Fe-Fe and Cr-Fe nearest neighbors,\cite{Ogura2011,Hasegawa1997,Lavrentiev2011} respectively.
They could be enough to maintain the ferromagnetic states at room temperatures.
However, practically, associated with the existence of magnetic dead layer in such system,\cite{NozakiPRA2016} Fe atom may diffuse into the Cr layer during the fabricating process at high temperatures.
Indeed, when exchanging the Fe and Cr atoms at the Cr/Fe interface in Fe 4ML, the total energy does not become so higher (50 meV/in-plane Fe). 

\begin{figure}
\centering
\includegraphics[width=75mm]{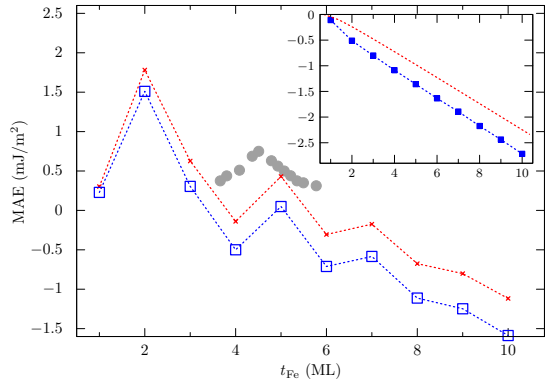}
\caption{Total magnetic anisotropy energy (MAE), compared with the experimental data (bullets)(ref. \onlinecite{NozakiPRA2016}). 
The open square symbol specifies the data estimated theoretically;   MAE(SOI)+MAE(MDI), and the cross data the sum of the theoretical MAE(SOI) and the shape anisotropy MAE(SA) estimated using the experimental magnetization. The inset shows the theoretical MAE(MDI) (filled squares) and experimental MAE(SA) (dots). The dashed lines are only for the guide of eyes.}
\label{MAEtotal}
\end{figure}

The shape anisotropy is realistically important for determining the whole magnetic anisotrpy. 
When we compare our MAE(SOI) with those from measurements, the shape anisotropy needs to be estimated precisely. 
 Here, we estimated the shape anisotropy energy, MAE(SA), in two ways; from the MDI\cite{Szunyogh1995} using the theoretical atomic magnetic moments and the estimation ($-\mu_{0}M_{s}^{2}/2$) using the experimental saturation magnetization ($M_{s}$).\cite{NozakiPRA2016} In the former, the 3MLs of Cr attached to the Fe layer were also taken into account. Figure \ref{MAEtotal} reports the total MAE, namely summation of MAE(SOI) and MAE(SA), in comparison with the available experimental data. This figure shows a good agreement with the experiment. As shown at the inset in Fig. \ref{MAEtotal}, the experimental MAE(SA) is much reduced in absolute, compared with the theoretical one, and gets a better agreement with the experiment on the total MAE. 
 
In the theoretical approach we obtained the magnetic dead layer (MDL) with 0.053 nm thick at most, while the experiment clearly showed the MDL with 0.1 nm thick.\cite{NozakiPRA2016}
The theoretical magnetization is larger than the experimental one by about 25 \% at the Fe 5 ML, as shown in the supplemental material (see Fig. S2).
These differences could be ascribed to an alloying effect at the Cr/Fe interface in the experiment. 
This allows to draw a suppression of the in-plane anisotropy from MDI contribution, or alternatively an enhancement in the strength of interface perpendicular anisotropy. 
The nearest neighbor pairs of Cr--Cr and Fe--Cr tends to enhances antiferromagnetic coupling between the magnetic moments, and possibly inducing a noncollinear magnetic structures due to a competition in magnetic couplings.\cite{Lavrentiev2011} Associated with such complexity in magnetic configuration, the MAE(SA) modulates due to the change in MDI. 

In summary, we performed the arguments on the vHS in the IRS of Fe/MgO interface and on the accuracy of MAE from the SOI, using the thickness dependences of the Fe layer with Cr underlayer. It was found that the DFT approach can describe the MAE with the numerical accuracy which is comparable to the experimental thickness dependence.  The vHS which appears at ${\bf k}_{1}$ in the band dispersions are found to play the important role in the perpendicular anisotropy. In the Fe 2ML, particularly, the flat bands consisting of $d_{xz}$, $d_{yz}$, $d_{xy}$, $d_{x^{2}-y^{2}}$ are located just below and above the Fermi level, which contribute to the large MAE of 2 mJ/m$^{2}$. This work shows that the potential of Fe/MgO over the previous works can be drawn by introducing the Cr underlayer, like a proximity effect, implying a possible materials design in the multi-functional interfaces through underlayer.


\begin{acknowledgments}

The authors thank T. Nozaki, Y. Suzuki, S. Yuasa, H. Sukegawa, M. Tsujikawa, Y. Miura, K. Nakamura, and M. Shirai for their stimulated discussions. The autor (T.O.) thanks A. Oshiyama for inspiring an aspect in the computational accuracies on MAE. The first-principles calculation was performed using the facilities of the Supercomputer Center, Institute for Solid State Physics, University of Tokyo, Japan. This research partly used computational resources of the K computer and other computers of the HPCI system provided by the AICS and Information Technology Center of Nagoya University through the HPCI System Research Project (Project ID:hp160227, hp160107). This work was partly supported by ImPACT Program of Council for Science, Technology and Innovation (Cabinet Office, Japan Government), and by the Computational Materials Science Initiative (CMSI), Japan, and by Kanazawa University SAKIGAKE Project. The author (T.O.) acknowledges the stimulated discussion in the meeting of the Cooperative Research Project of the Research Institute of Electrical Communication, Tohoku University (Project No. H25/B04). The author (M.O.) acknowledges JSPS Research Fellowships for Young Scientists (Grant No. 27-4610) and the authors (N.I. and I.P.) acknowledges Japanese Government (Monbukagakusho: MEXT) Scholarship in the Program for the Development of Global Human Resources. 
\end{acknowledgments}

\end{document}


\begin{center}

\large{{\Large Supplementary Materials for}\\ 
``Reinvestigation on large perpendicular magnetic anisotroy in Fe/MgO interface from first-principles approach''}\\[1cm]

Nurul Ikhsan,$^{1,2}$ Tomosato Kanagawa,$^{1}$ Indra Pardede,$^{1,3}$\\
Masao Obata,$^{1,4}$ Tatsuki Oda$^{1,4}$\\[1cm]

$^{1}$Graduate School of Natural Science and Technology, Kanazawa University, Kakuma, Kanazawa 920-1192, Japan\\
$^{2}$School of Computing, Telkom University, Bandung 40257, Indonesia\\
$^{3}$Department of Physics, Institut Teknologi Sumatera, Lampung Selatan 35365, Indonesia\\
$^{4}$Institute of Science and Engineering, Kanazawa University, Kakuma, Kanazawa 920-1192, Japan
\end{center}

\newpage

\begin{figure}[t]
\begin{center}
\includegraphics[width=75mm]{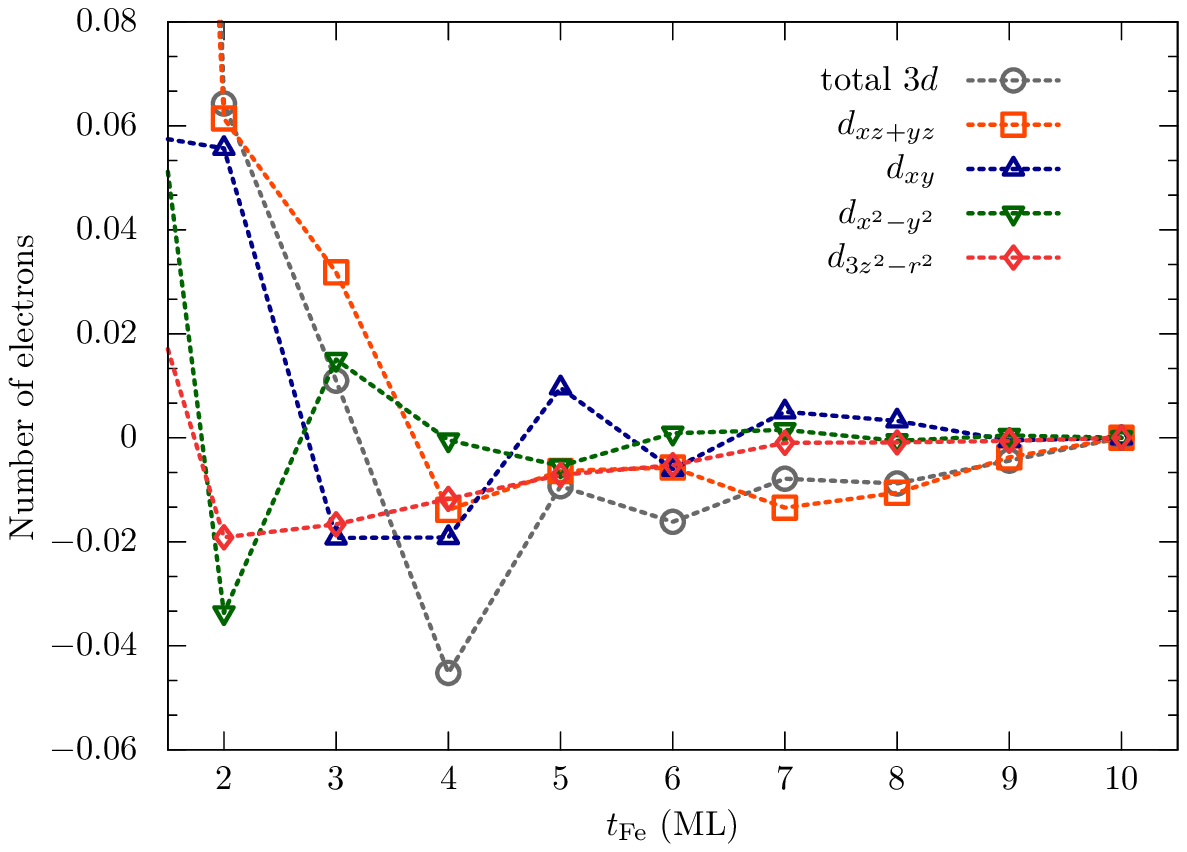}
\end{center}
\hspace*{-5mm}
FIG. S1: Thickness dependence of the number of 3d electron on the interface Fe. The relative values with respect to those of Fe 10ML are shown. The colors of grey, orange (yellow), blue, green, and red are used for total 3$d$ orbital, $d_{xz+yz}$, $d_{xy}$, $d_{x^{2}-y^{2}}$, and $d_{3z^{2}-r^{2}}$, respectively.
\label{SFigNOE}
\end{figure}

\begin{figure}[t]
\begin{center}
\includegraphics[width=75mm]{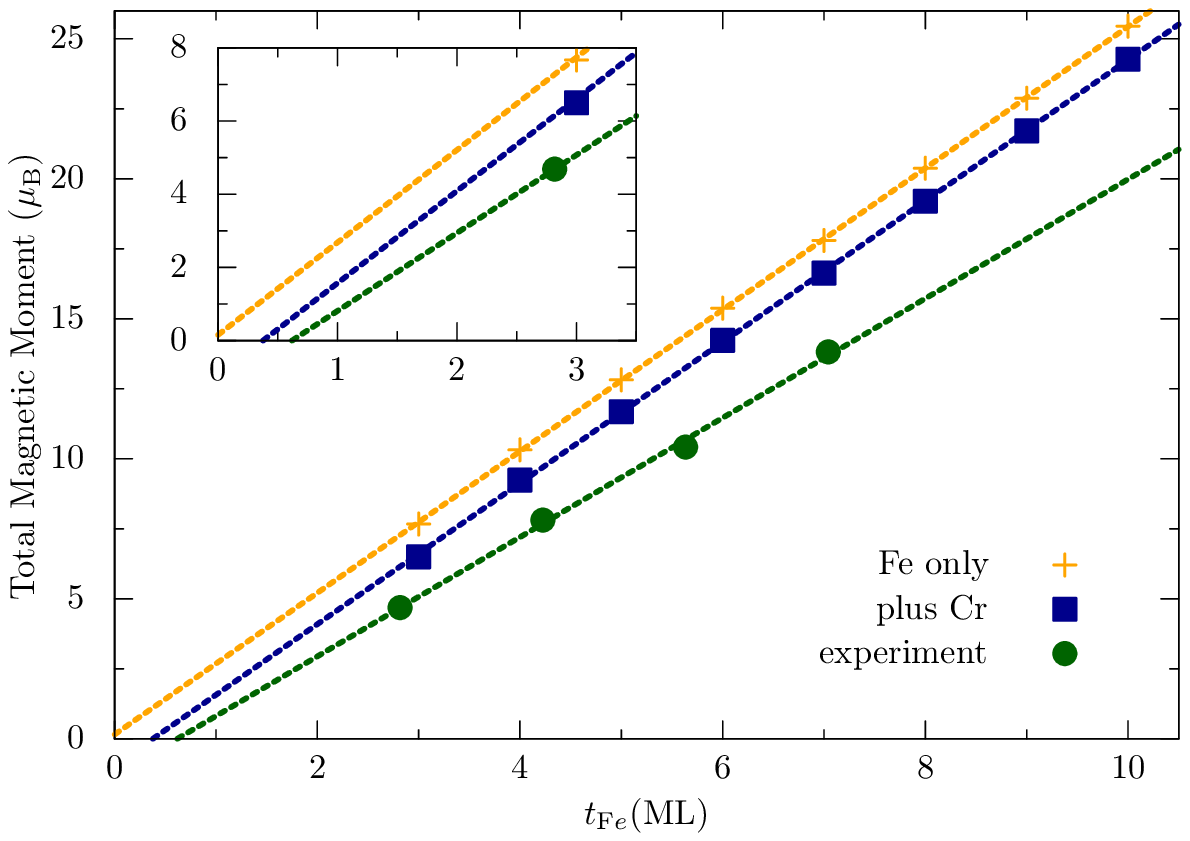}
\end{center}
\hspace*{-5mm}
FIG. S2: Thickness dependence of magnetizations, compared with the available experimental data(ref. \onlinecite{NozakiPRA2016}). Plus symbol indicates total magnetizations contributed from Fe atom only, solid box includes the contribution of Cr, bullet indicates experimental data.
\label{SFigmag}
\end{figure}